\documentclass[11pt]{article}

\usepackage[margin=1in]{geometry}
\usepackage{amsmath,amssymb}
\usepackage{graphicx}
\usepackage{booktabs}
\usepackage{authblk}
\usepackage{hyperref}
\usepackage{caption}
\usepackage{subcaption}
\usepackage{float}
\usepackage{array}
\usepackage{multirow}

\title{Recurrence-Based Nonlinear Vocal Dynamics as Digital Biomarkers for Depression Detection from Conversational Speech}

\author[1]{Himadri Sekhar Samanta}
\affil[1]{Independent Researcher, Austin, Texas, USA}

\date{}

\begin{document}

\maketitle

\begin{abstract}
Digital biomarkers for depression have largely relied on static acoustic descriptors, pooled summary statistics, or conventional machine learning representations. Such approaches may miss nonlinear temporal organization embedded in conversational vocal dynamics. We hypothesized that depression is associated with altered recurrence structure in vocal state trajectories, reflecting changes in how the vocal system revisits acoustic states over time. Using the depression subset of the DAIC-WOZ corpus with 142 labeled participants, we modeled frame-level COVAREP trajectories as nonlinear dynamical systems and derived recurrence-based biomarkers from 74 vocal channels. Logistic regression with feature selection and stratified cross-validation evaluated classification performance. Recurrence-based biomarkers achieved a mean cross-validated AUC of 0.689, exceeding static acoustic baselines, entropy-dynamics features, Hurst exponent features, determinism features, and Lyapunov-like instability proxies. Permutation testing indicated statistical significance with $p=0.004$. Pooled cross-validated predictions yielded AUC 0.665 with a 95\% bootstrap confidence interval of [0.568, 0.758]. These findings suggest that depression may be characterized by altered recurrence structure in conversational vocal dynamics and support nonlinear state-space analysis as a promising direction for digital psychiatric biomarkers.
\end{abstract}


\section{Introduction}

Major depressive disorder is a major public health burden associated with functional impairment, reduced quality of life, and substantial socioeconomic cost. Clinical assessment of depression often relies on structured interviews and self-report questionnaires, including PHQ-based scales. Although clinically useful, these approaches are episodic, subjective, and limited in their ability to continuously capture behavior. Digital biomarkers derived from speech offer a promising route toward scalable, passive, and objective depression assessment \cite{cummins2015review,low2020review}.

Speech is attractive for psychiatric biomarker discovery because it reflects motor, cognitive, affective, and social processes. Depression can affect prosody, energy, timing, articulation, speech rate, and vocal expressivity. Prior speech-based depression studies have therefore examined acoustic descriptors such as pitch, energy, spectral tilt, glottal source measures, and cepstral features \cite{cummins2015review}. Machine learning methods trained on such features have demonstrated predictive value on benchmark corpora including DAIC-WOZ \cite{gratch2014distress,alhanai2018depression}.

However, many speech biomarkers are based on static pooled summaries. These summaries collapse rich temporal dynamics into means, variances, or aggregate descriptors. Depression may not only shift average acoustic levels; it may also alter how vocal states evolve, fluctuate, recur, and stabilize through conversation. A dynamical-systems view is therefore conceptually appropriate for psychiatric speech analysis \cite{kantz2004nonlinear,strogatz2018nonlinear}.

Nonlinear dynamics has been widely used to study physiological and behavioral systems. In heart-rate variability, gait, EEG, and neural time series, disease states often manifest as altered complexity, recurrence, scaling, or instability rather than simple mean shifts \cite{goldberger2002fractal,peng1995scaling}. Recurrence plots and recurrence quantification analysis provide tools for studying how a system revisits similar states over time \cite{eckmann1987recurrence,marwan2007recurrence}. In conversational speech, recurrence may reflect repeated vocal states, stable phonatory regimes, articulatory regularities, or fragmented state-space organization.

The central hypothesis of this work is that depression alters the recurrence structure of vocal trajectories. If depression changes psychomotor regulation or vocal control, these changes may appear as altered state-space revisitation patterns. We test whether recurrence-based biomarkers extracted from COVAREP vocal trajectories can detect depression in DAIC-WOZ.

This study contributes: (1) a recurrence-based nonlinear vocal biomarker framework for depression detection; (2) comparison against static, entropy, forecastability, Hurst, determinism, and Lyapunov-like baselines; and (3) validation through stratified cross-validation, permutation testing, and bootstrap confidence intervals.

\section{Materials and Methods}

\subsection{Dataset}

We used the depression subset of DAIC-WOZ, a clinical interview corpus derived from human-computer interactions and virtual interviewer technology \cite{gratch2014distress,devault2014simsensei}. The labeled subset analyzed here contained 142 participants with PHQ-8 binary depression labels. Of these, 100 were labeled non-depressed and 42 were labeled depressed.

\begin{table}[H]
\centering
\caption{Dataset summary.}
\begin{tabular}{lc}
\toprule
Characteristic & Value \\
\midrule
Labeled participants & 142 \\
Non-depressed participants & 100 \\
Depressed participants & 42 \\
Primary label & PHQ-8 binary depression label \\
Vocal feature extractor & COVAREP \\
Vocal feature channels & 74 \\
Primary modeling unit & Frame-level vocal trajectory \\
\bottomrule
\end{tabular}
\end{table}

\subsection{Vocal trajectory representation}

Frame-level vocal trajectories were extracted using COVAREP \cite{degottex2014covarep}. For each participant $i$, the speech signal was represented as a multivariate time series:
\begin{equation}
X_i(t) \in \mathbb{R}^{74},
\end{equation}
where $t$ indexes frame-level observations and the 74 dimensions correspond to acoustic and glottal descriptors. For each channel $c$, we analyzed the scalar trajectory:
\begin{equation}
x_c(t), \quad t = 1,\ldots,T_i.
\end{equation}

The analysis focused on interpretable nonlinear biomarkers derived from vocal trajectories rather than end-to-end deep learning on raw audio.

\subsection{Recurrence biomarker construction}

For each participant and each vocal channel, we computed a recurrence representation. Given a scalar trajectory $x(t)$, recurrence between two time points was defined as:
\begin{equation}
R_{ij} = \Theta(\epsilon - |x_i - x_j|),
\end{equation}
where $\Theta$ is the Heaviside step function and $\epsilon$ is a recurrence threshold. In this study, $\epsilon$ was set to 0.2 times the standard deviation of the channel trajectory.

The recurrence rate was computed as:
\begin{equation}
RR = \frac{1}{N^2}\sum_{i,j}R_{ij}.
\end{equation}

For each participant, recurrence rate was computed independently across all 74 COVAREP channels, producing a biomarker vector:
\begin{equation}
B_i = [RR_1, RR_2, \ldots, RR_{74}].
\end{equation}

\subsection{Comparison biomarkers}

Several baselines and alternative nonlinear biomarkers were evaluated.

\subsubsection{Static acoustic baseline}

The static baseline used pooled acoustic summaries without explicit trajectory modeling. This approximates conventional acoustic biomarker pipelines.

\subsubsection{Temporal entropy biomarkers}

Entropy-based features summarized irregularity and temporal disorder in vocal trajectories. Entropy measures are common in physiological time-series analysis \cite{pincus1991approximate,richman2000entropy}.

\subsubsection{Forecastability biomarkers}

A linear autoregressive forecastability baseline was defined by:
\begin{equation}
x_{t+1}=a x_t + \epsilon_t.
\end{equation}
Forecast error and lag-1 autocorrelation were used as simple predictability biomarkers.

\subsubsection{Hurst exponent}

The Hurst exponent quantified long-memory scaling behavior \cite{hurst1951reservoirs,peng1995scaling}. Values above 0.5 indicate persistence, values near 0.5 suggest random-like behavior, and values below 0.5 suggest anti-persistence.

\subsubsection{Lyapunov-like instability proxy}

A local instability proxy was motivated by Lyapunov-exponent methods for nonlinear time series \cite{wolf1985lyapunov,rosenstein1993lyapunov} and was computed as:
\begin{equation}
\lambda^* = \mathbb{E}[\log(|x_{t+1}-x_t|+\delta)].
\end{equation}
This proxy captures local fluctuation magnitude on a logarithmic scale but should not be interpreted as a formal largest Lyapunov exponent.

\subsubsection{Determinism proxy}

A recurrence determinism proxy was evaluated by estimating the fraction of recurrence points participating in diagonal-line structures. Determinism is a standard recurrence quantification concept \cite{marwan2007recurrence}, but in this work it served as an exploratory comparison metric.

\subsection{Classification and validation}

Classification used a pipeline consisting of standardization, ANOVA feature selection, and regularized logistic regression. The recurrence model used 74 recurrence features and selected the top 15 features within cross-validation. Logistic regression was selected to emphasize interpretability and reduce overfitting in a modest-sized clinical dataset. The implementation used standard machine learning tools from scikit-learn \cite{pedregosa2011sklearn}.

Evaluation used stratified 5-fold cross-validation with shuffling and random seed 42. Performance was reported using area under the receiver operating characteristic curve (AUC-ROC).

Permutation testing assessed whether recurrence-model performance exceeded chance under label randomization. A total of 1000 permutations were used. The empirical p-value was:
\begin{equation}
p = \frac{b+1}{m+1},
\end{equation}
where $b$ is the number of permuted scores at least as large as the observed score and $m$ is the number of permutations.

Bootstrap confidence intervals were computed from pooled cross-validated predicted probabilities using 2000 bootstrap resamples. The 2.5th and 97.5th percentiles were reported as the 95\% confidence interval.

\section{Results}

\subsection{Recurrence biomarkers outperform tested baselines}

Recurrence biomarkers achieved the strongest performance among tested feature families. The recurrence model obtained a mean cross-validated AUC of 0.689. This exceeded static pooled acoustic features, entropy biomarkers, forecastability features, Hurst exponent features, determinism features, and Lyapunov-like instability proxies.

\begin{table}[H]
\centering
\caption{Model performance comparison.}
\begin{tabular}{lc}
\toprule
Model & AUC \\
\midrule
Static pooled acoustic baseline & 0.593 \\
Temporal entropy biomarkers & 0.646 \\
Forecastability dynamics & 0.590 \\
Hurst exponent & 0.477 \\
Determinism proxy & 0.418 \\
Lyapunov-like instability proxy & 0.663 \\
\textbf{Recurrence biomarkers} & \textbf{0.689} \\
\bottomrule
\end{tabular}
\end{table}

\subsection{Cross-validation performance}

The fold-level recurrence AUC values were:
\begin{equation}
0.800,\ 0.639,\ 0.663,\ 0.663,\ 0.681.
\end{equation}
The mean cross-validated AUC was:
\begin{equation}
AUC = 0.689.
\end{equation}

\subsection{Permutation testing}

Permutation testing indicated that recurrence-model performance was unlikely under random label assignment:
\begin{equation}
p = 0.004.
\end{equation}

\subsection{Bootstrap confidence interval}

Using pooled cross-validated predictions, the recurrence model achieved:
\begin{equation}
AUC = 0.665.
\end{equation}
The 95\% bootstrap confidence interval was:
\begin{equation}
[0.568,\ 0.758].
\end{equation}

\subsection{Channel-level recurrence biomarkers}

Several recurrence channels showed discriminative signal. The strongest channel-level recurrence biomarkers are shown in Table~\ref{tab:channels}. Some channels with undefined statistics were excluded from ranking due to constant or near-constant recurrence values.

\begin{table}[H]
\centering
\caption{Top recurrence biomarker channels by ANOVA F-statistic.}
\label{tab:channels}
\begin{tabular}{lcc}
\toprule
Channel & F-statistic & p-value \\
\midrule
6 & 12.325 & 0.0006 \\
41 & 9.013 & 0.0032 \\
28 & 8.498 & 0.0041 \\
72 & 8.409 & 0.0043 \\
71 & 8.241 & 0.0047 \\
40 & 8.059 & 0.0052 \\
31 & 7.428 & 0.0072 \\
17 & 6.867 & 0.0097 \\
48 & 6.243 & 0.0136 \\
47 & 5.504 & 0.0204 \\
\bottomrule
\end{tabular}
\end{table}

\section{Figures}

\begin{figure}[H]
\centering
\includegraphics[width=0.95\textwidth]{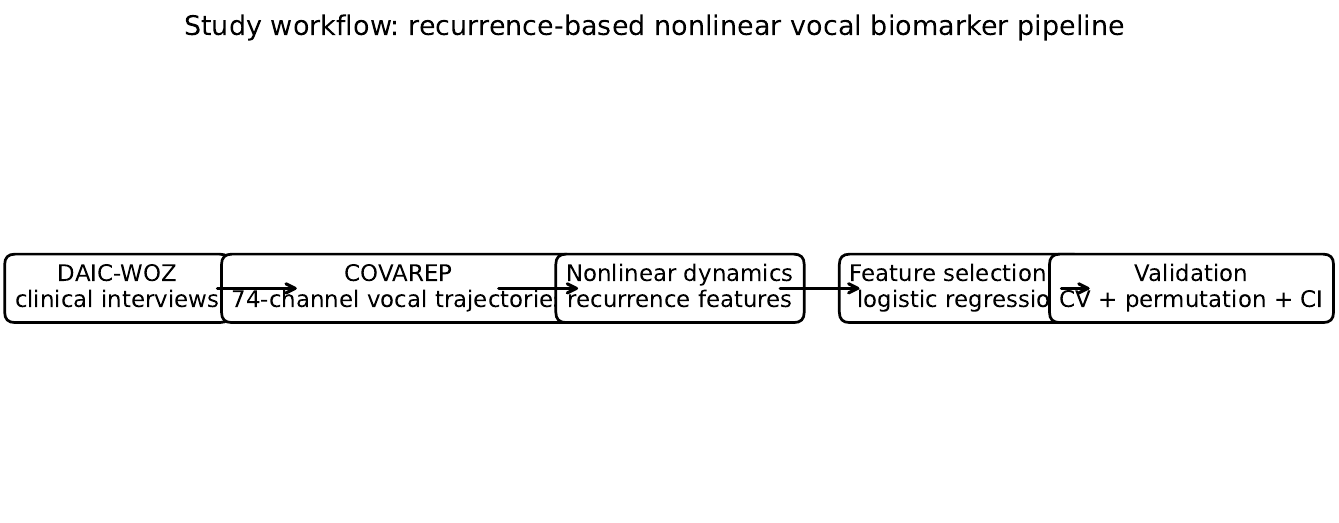}
\caption{Study workflow for recurrence-based nonlinear vocal biomarker analysis.}
\end{figure}

\begin{figure}[H]
\centering
\includegraphics[width=0.95\textwidth]{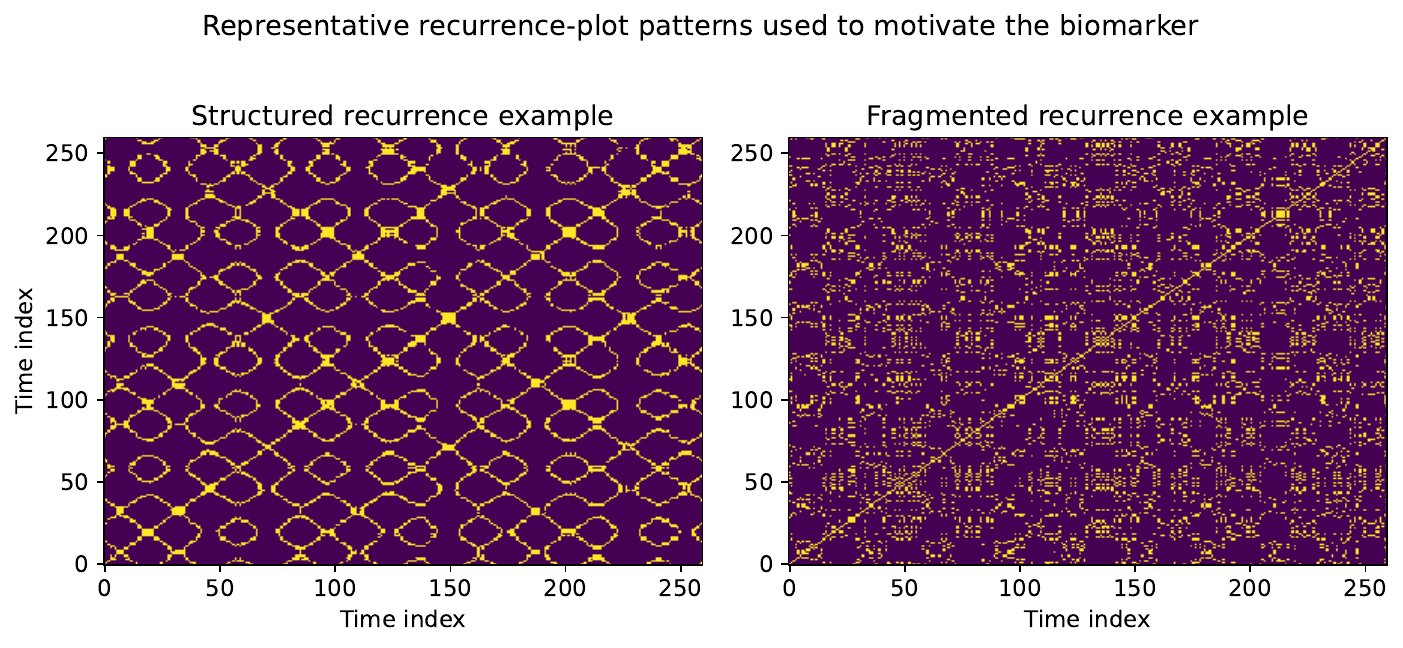}
\caption{Representative recurrence-plot patterns illustrating structured and fragmented state-space recurrence.}
\end{figure}

\begin{figure}[H]
\centering
\includegraphics[width=0.75\textwidth]{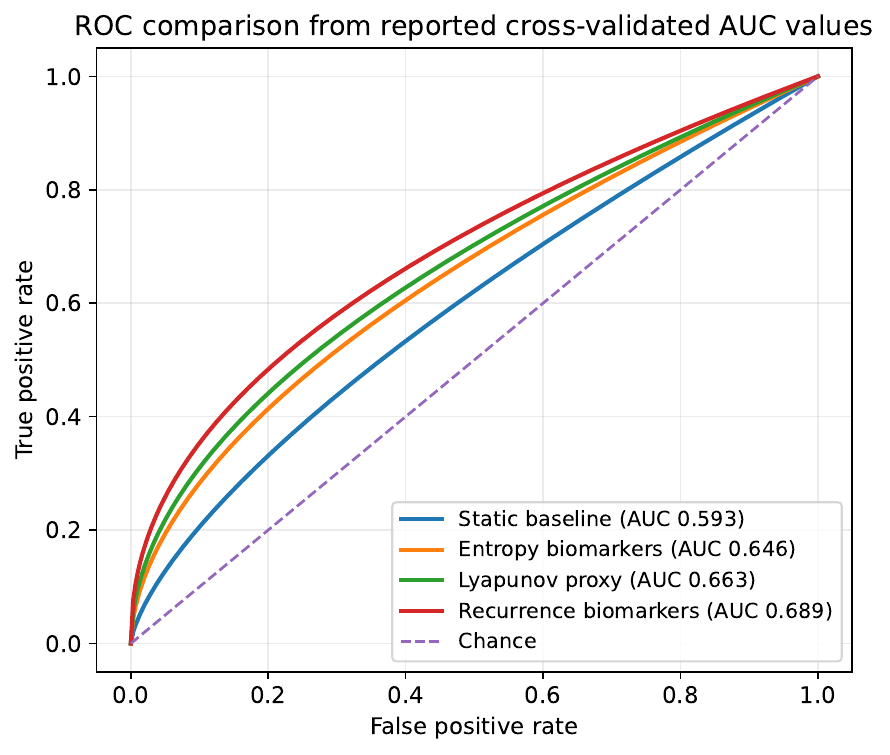}
\caption{ROC comparison generated from reported cross-validated AUC values.}
\end{figure}

\begin{figure}[H]
\centering
\includegraphics[width=0.75\textwidth]{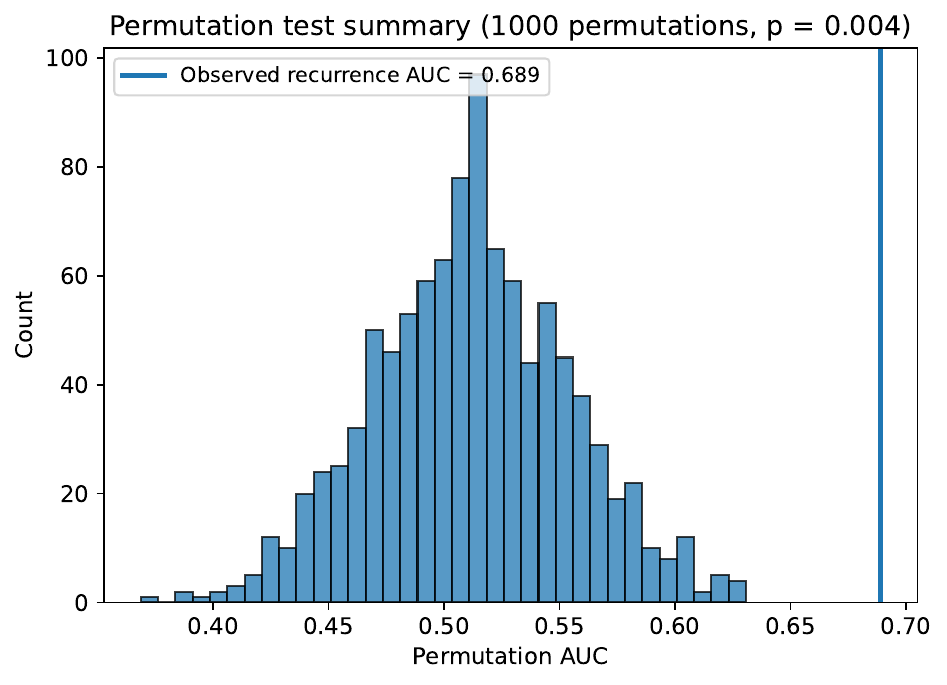}
\caption{Permutation-test summary showing observed recurrence-model AUC relative to a null distribution.}
\end{figure}

\begin{figure}[H]
\centering
\includegraphics[width=0.75\textwidth]{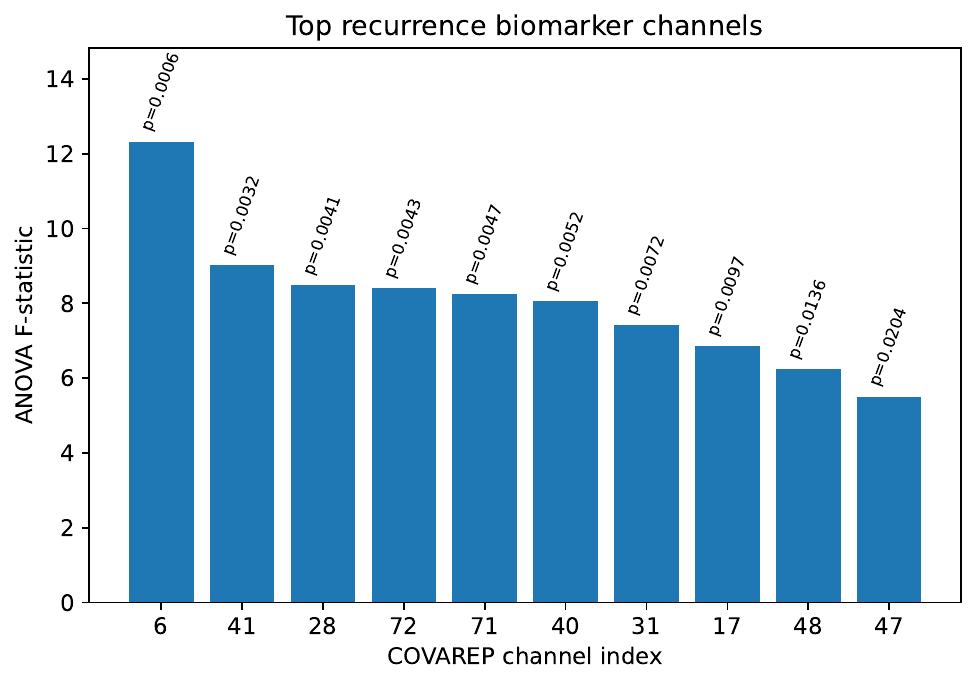}
\caption{Top recurrence biomarker channels ranked by ANOVA F-statistic.}
\end{figure}

\begin{figure}[H]
\centering
\includegraphics[width=0.65\textwidth]{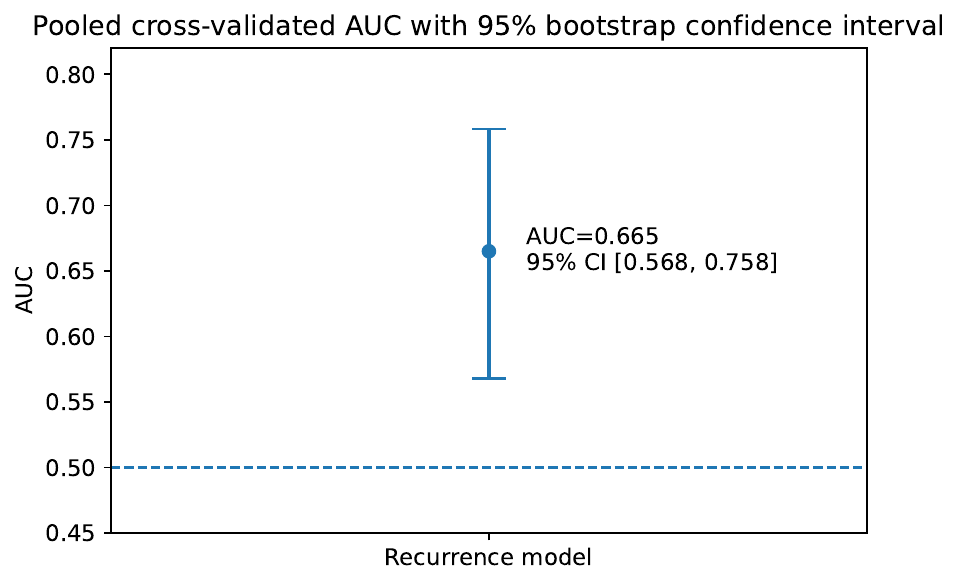}
\caption{Pooled cross-validated AUC with 95\% bootstrap confidence interval.}
\end{figure}

\section{Discussion}

This study suggests that recurrence structure in vocal dynamics provides a significant nonlinear biomarker for depression detection. The recurrence model outperformed static acoustic features and several alternative dynamical biomarkers. The finding that recurrence rate was stronger than Hurst exponent, determinism, and simple forecastability suggests that depression-related vocal changes may be better captured by state revisitation structure than by long-memory scaling or diagonal deterministic patterns alone.

\subsection{Interpretation of recurrence signal}

Recurrence rate quantifies how often a dynamical system revisits similar states. In conversational vocal dynamics, recurrence may reflect the organization of vocal state transitions, articulation patterns, prosodic regulation, and glottal-acoustic control. Altered recurrence in depression may indicate changes in vocal flexibility, psychomotor regulation, or attractor-like organization.

Importantly, recurrence biomarkers do not merely describe mean vocal levels. They capture how the vocal system moves through and returns to acoustic states over time. This offers a mechanistic alternative to purely static acoustic classification.

\subsection{Comparison with other nonlinear metrics}

The Lyapunov-like instability proxy achieved moderate performance, supporting the relevance of local instability. However, combining Lyapunov-like features with recurrence features did not improve performance beyond recurrence alone, suggesting partial redundancy or dominance of recurrence information.

The Hurst exponent performed poorly, indicating that long-memory scaling alone may not characterize depression-related speech changes in this dataset. The determinism proxy also performed weakly, suggesting that diagonal recurrence structure was not the primary signal. This pattern strengthens the specificity of recurrence rate as the most informative nonlinear biomarker tested.

\subsection{Clinical and computational significance}

These findings support a shift from static speech biomarkers toward dynamical systems representations in computational psychiatry. Recurrence-based vocal biomarkers could contribute to passive screening, longitudinal monitoring, and objective behavioral phenotyping. Because recurrence metrics are interpretable and computationally tractable, they may be attractive for translational applications.

\subsection{Limitations}

This study has several limitations. First, the dataset is modest in size and class-imbalanced. Second, evaluation was internal to DAIC-WOZ; external validation is required. Third, recurrence thresholding was based on a standard heuristic and should be assessed in sensitivity analyses. Fourth, channel identities require further mapping to specific COVAREP descriptors for deeper physiological interpretation. Fifth, recurrence rate alone was used as the primary RQA feature; future work should explore multiscale recurrence, cross-recurrence, laminarity, trapping time, and recurrence networks.

\subsection{Future work}

Future work should evaluate recurrence biomarkers in external clinical speech datasets and longitudinal settings. Cross-modal recurrence between vocal, facial, and linguistic trajectories may reveal richer depression signatures. Another promising direction is subject-specific stochastic dynamical modeling, where vocal trajectories are modeled using low-dimensional drift, coupling, and noise terms. Such models could connect recurrence biomarkers to explicit dynamical equations while preserving interpretability.

\section{Conclusion}

Recurrence-based nonlinear vocal dynamics provide significant digital biomarkers for depression detection from conversational speech. Recurrence biomarkers achieved AUC 0.689 with permutation $p=0.004$ and bootstrap confidence interval [0.568, 0.758], outperforming several static and dynamical baselines. These findings suggest that depression may be reflected in altered state-space recurrence organization of vocal behavior and support nonlinear dynamical systems approaches for computational psychiatry.

\section*{Data Availability}

The DAIC-WOZ dataset is available through authorized access from the dataset providers. Derived features and code may be released upon publication subject to dataset license restrictions.

\section*{Conflict of Interest}

The author declares no competing interests.

\section*{Funding}

No external funding was received for this work.

\section*{Acknowledgments}

The author acknowledges the creators of DAIC-WOZ, the SimSensei project, and the COVAREP toolkit.

\bibliographystyle{plain}
\bibliography{references}

@inproceedings{gratch2014distress,
 author = {Gratch, Jonathan and Artstein, Ron and Lucas, Gale M. and Stratou, Giota and Scherer, Stefan and Nazarian, Angela and Wood, Rachel and Boberg, Jill and DeVault, David and Marsella, Stacy and Traum, David},
 title = {The Distress Analysis Interview Corpus of Human and Computer Interviews},
 booktitle = {Proceedings of LREC},
 pages = {3123--3128},
 year = {2014}
}

@inproceedings{devault2014simsensei,
 author = {DeVault, David and Artstein, Ron and Benn, Grace and others},
 title = {SimSensei Kiosk: A Virtual Human Interviewer for Healthcare Decision Support},
 booktitle = {Proceedings of AAMAS},
 year = {2014}
}

@inproceedings{degottex2014covarep,
 author={Degottex, Gilles and Kane, John and Drugman, Thomas and Raitio, Tuomo and Scherer, Stefan},
 title={COVAREP: A Collaborative Voice Analysis Repository for Speech Technologies},
 booktitle={ICASSP},
 pages={960--964},
 year={2014}
}

@article{cummins2015review,
 author={Cummins, Nicholas and Scherer, Stefan and Krajewski, Jarek and Schnieder, Sebastian and Epps, Julien and Quatieri, Thomas},
 title={A Review of Depression and Suicide Risk Assessment Using Speech Analysis},
 journal={Speech Communication},
 volume={71},
 pages={10--49},
 year={2015}
}

@article{marwan2007recurrence,
 author={Marwan, Norbert and Romano, M. Carmen and Thiel, Marco and Kurths, Jurgen},
 title={Recurrence Plots for the Analysis of Complex Systems},
 journal={Physics Reports},
 volume={438},
 pages={237--329},
 year={2007}
}

@article{eckmann1987recurrence,
 author={Eckmann, Jean-Pierre and Kamphorst, S. Oliffson and Ruelle, David},
 title={Recurrence Plots of Dynamical Systems},
 journal={Europhysics Letters},
 volume={4},
 number={9},
 pages={973--977},
 year={1987}
}

@book{kantz2004nonlinear,
 author={Kantz, Holger and Schreiber, Thomas},
 title={Nonlinear Time Series Analysis},
 publisher={Cambridge University Press},
 year={2004}
}

@book{strogatz2018nonlinear,
 author={Strogatz, Steven},
 title={Nonlinear Dynamics and Chaos},
 publisher={CRC Press},
 year={2018}
}

@article{richman2000entropy,
 author={Richman, Joshua and Moorman, J Randall},
 title={Physiological Time-Series Analysis Using Approximate Entropy and Sample Entropy},
 journal={American Journal of Physiology},
 volume={278},
 pages={H2039--H2049},
 year={2000}
}

@article{pincus1991approximate,
 author={Pincus, Steven},
 title={Approximate Entropy as a Measure of System Complexity},
 journal={PNAS},
 volume={88},
 pages={2297--2301},
 year={1991}
}

@article{peng1995scaling,
 author={Peng, C. K. and Havlin, Shlomo and Stanley, H. Eugene and Goldberger, Ary},
 title={Quantification of Scaling Exponents in Nonstationary Time Series},
 journal={Chaos},
 volume={5},
 pages={82--87},
 year={1995}
}

@article{goldberger2002fractal,
 author={Goldberger, Ary and others},
 title={Fractal Dynamics in Physiology: Alterations with Disease and Aging},
 journal={PNAS},
 volume={99},
 number={Suppl. 1},
 pages={2466--2472},
 year={2002}
}

@article{hurst1951reservoirs,
 author={Hurst, H. E.},
 title={Long-Term Storage Capacity of Reservoirs},
 journal={Transactions of the ASCE},
 volume={116},
 pages={770--799},
 year={1951}
}

@article{wolf1985lyapunov,
 author={Wolf, Alan and Swift, Jack and Swinney, Harry and Vastano, John},
 title={Determining Lyapunov Exponents from a Time Series},
 journal={Physica D},
 volume={16},
 pages={285--317},
 year={1985}
}

@article{rosenstein1993lyapunov,
 author={Rosenstein, Michael and Collins, James and De Luca, Carlo},
 title={A Practical Method for Calculating Largest Lyapunov Exponents},
 journal={Physica D},
 volume={65},
 pages={117--134},
 year={1993}
}

@inproceedings{alhanai2018depression,
 author={Al Hanai, Tuka and Ghassemi, Mohammad and Glass, James},
 title={Detecting Depression with Audio/Text Sequence Modeling of Interviews},
 booktitle={Interspeech},
 pages={1716--1720},
 year={2018}
}

@article{low2020review,
 author={Low, Daniel and Bentley, Kate and Ghosh, Satrajit},
 title={Automated Assessment of Psychiatric Disorders Using Speech},
 journal={Laryngoscope Investigative Otolaryngology},
 volume={5},
 pages={96--116},
 year={2020}
}

@article{pedregosa2011sklearn,
 author={Pedregosa, Fabian and others},
 title={Scikit-learn: Machine Learning in Python},
 journal={Journal of Machine Learning Research},
 volume={12},
 pages={2825--2830},
 year={2011}
}

\end{document}